\documentclass[conference]{IEEEtran}

\usepackage[utf8]{inputenc}
\usepackage{graphicx}
\usepackage{xcolor}  
\usepackage{amsmath,amssymb}
\usepackage{ifthen}
\usepackage{xspace}
\usepackage{xspace}
\usepackage{multicol}
\usepackage{enumitem}
\usepackage[ruled,vlined, linesnumbered]{algorithm2e}
\usepackage{color}
\usepackage{listings}
\usepackage{pifont}
\usepackage{multirow}
\usepackage{url}
\usepackage{hyperref}
\usepackage{cleveref}
\usepackage{amsthm}
\usepackage{array}
\usepackage{float}
\usepackage{booktabs}
\usepackage{subcaption}
\usepackage[backend=bibtex,sorting=none,style=ieee,url=false,isbn=false]{biblatex}

\newcommand{\enumA}{(i)}
\newcommand{\enumB}{(ii)}
\newcommand{\enumC}{(iii)}
\newcommand{\enumD}{(iv)}

\crefname{section}{\S}{\S\S}
\crefname{subsection}{\S}{\S\S}
\crefname{subsubsection}{\S}{\S\S}
\crefname{table}{Table}{Tables}
\crefname{figure}{Fig.}{Figs.}

\newcommand{\mypar}[1]{\noindent\textbf{#1.}\xspace}

\providecommand{\subparagraph}{}
\usepackage{titlesec}
\titlespacing*{\section}{0pt}{0.1\baselineskip}{0.1\baselineskip}
\titlespacing*{\subsection}{0pt}{0.1\baselineskip}{0.1\baselineskip}
\ifCLASSINFOpdf
  \else
  \fi

\newcommand{\peas}{\textsc{PEAS}\xspace}
\newcommand{\xsearch}{\textsc{X-Search}\xspace}
\newcommand{\tor}{\textsc{Tor}\xspace}

\newcommand{\tmn}{\textsc{TrackMeNot}\xspace}
\newcommand{\tmns}{\textsc{TMN}\xspace}
\newcommand{\goopir}{\textsc{GooPIR}\xspace}

\newcommand{\name}{\textsc{Cyclosa}\xspace}
\newcommand{\SYS}{\textsc{Cyclosa}\xspace}

\newcommand{\y}{$\checkmark$}
\newcommand{\n}{$\times$}

\bibliography{biblio}

\renewbibmacro{finentry}{    \iffieldequalstr{entrykey}{brenner2016securekeeper}    {\finentry\newpage}
    {\finentry}}

\usepackage{tikz}
\newcommand\copyrighttext{  \footnotesize \textcopyright 2018 IEEE.
    Personal use of this material is permitted.
    Permission from IEEE must be obtained for all other uses,
    in any current or future media, including reprinting/republishing this
    material for advertising or promotional purposes, creating new collective
    works, for resale or redistribution to servers or
    lists, or reuse of any copyrighted component of this work in other works.
    Pre-print version. Presented in the \href{http://icdcs2018.ocg.at}{38th IEEE International Conference on Distributed Computing Systems (ICDCS '18)}. For the final published version, refer to DOI \href{https://doi.org/10.1109/ICDCS.2018.00053}{10.1109/ICDCS.2018.00053}}
\newcommand\copyrightnotice{\begin{tikzpicture}[remember picture,overlay]
\node[anchor=south,yshift=10pt,fill=yellow!20] at (current page.south) {\fbox{\parbox{\dimexpr\textwidth-\fboxsep-\fboxrule\relax}{\copyrighttext}}};
\end{tikzpicture}}
\begin{document}

\title{\SYS: Decentralizing Private Web Search\\Through SGX-Based Browser Extensions}

\author{
\\[-5.0ex]

\IEEEauthorblockN{
Rafael Pires\IEEEauthorrefmark{2},
David Goltzsche\IEEEauthorrefmark{3},
Sonia Ben Mokhtar\IEEEauthorrefmark{1},
Sara Bouchenak\IEEEauthorrefmark{1},\\
Antoine Boutet\IEEEauthorrefmark{4}, 
Pascal Felber\IEEEauthorrefmark{2},
R\"udiger Kapitza\IEEEauthorrefmark{3},
Marcelo Pasin\IEEEauthorrefmark{2} and
Valerio Schiavoni\IEEEauthorrefmark{2}
}

\IEEEauthorblockA{
\IEEEauthorrefmark{1}INSA Lyon, CNRS, LIRIS, Lyon, France, 
\{firstname.lastname\}@insa-lyon.fr
}

\IEEEauthorblockA{
\IEEEauthorrefmark{2}University of Neuch\^atel, Switzerland, 
\{firstname.lastname\}@unine.ch}

\IEEEauthorblockA{
\IEEEauthorrefmark{3}TU Braunschweig, Germany, 
\{lastname\}@ibr.cs.tu-bs.de}

\IEEEauthorblockA{
\IEEEauthorrefmark{4}Univ Lyon, INSA Lyon, Inria, CITI, F-69621 Villeurbanne, France,
\{firstname.lastname\}@insa-lyon.fr
}

\\[-5.0ex]
}

\maketitle
\copyrightnotice

\begin{abstract}

By regularly querying Web search engines, users (unconsciously) disclose large amounts of their personal data as part of their search queries, among which some might reveal sensitive information (e.g. health issues, sexual, political or religious preferences).
Several solutions exist to allow users querying search engines while improving privacy protection. 
However, these solutions suffer from a number of limitations: some are subject to user re-identification attacks, while others lack scalability or are unable to provide accurate results.

This paper presents \name{}, a secure, scalable and accurate private Web search solution. 
\name{} improves security by relying on trusted execution environments~(TEEs) as provided by Intel SGX.
Further, \name{} proposes a novel adaptive privacy protection solution that reduces the risk of user re-identification. 
\name{} sends fake queries to the search engine and dynamically adapts their count according to the sensitivity of the user query. 
In addition, \name{} meets scalability as it is fully decentralized, spreading the load for distributing fake queries among other nodes.
Finally, \name{} achieves accuracy of Web search as it handles the real query and the fake queries separately, in contrast to other existing solutions that mix fake and real query results.

\end{abstract}

\IEEEpeerreviewmaketitle

\section{Introduction}
\label{sec:introduction}

Search engines have become an essential service for finding content on the Internet. However, by regularly querying these services, users disclose
large amounts of their personal data, comprising privacy-sensitive information such as their health status, religious, political or sexual preferences as shown during the AOL scandal when this search engine released a pseudomized dataset of search queries~\cite{aol-leak-scary-1,aol-leak-scary-2}. 

To limit the disclosure of personal information, many private Web search solutions have been proposed in the last decade. 
They can be classified in two categories. The first category of solutions enforce \textit{unlinkability} between users and their queries by hiding users' identity through anonymous communication~\cite{dingledine2004tor,corrigan2010dissent,mokhtar2013rac}.

However, studies have shown that anonymously sending queries to the search engine is not sufficient to actually protect users' privacy~\cite{peddinti2014web,petit2016simattack}. 
Indeed, a search engine that has prior knowledge about users (e.g.,~user profiles built from past user queries) can link back a large proportion of anonymous search queries to their originating user by running \emph{re-identification attacks}.
In order to mitigate this risk, a second category of solutions have been proposed.  
These solutions enforce \textit{indistinguishability} of the user interests by sending fake queries on behalf of the user~\cite{howe2009trackmenot,domingo2009h}.
Other solutions combine unlinkability and indistinguishability~\cite{petit2015peas,benmokhtar:hal-01588883}.
Additionally, existing privacy protection solutions suffer from the following limitations:
\enumA~their scalability is limited;
\enumB~their protection level is set in a static way; and
\enumC~their Web responses are inaccurate.
In fact, relying on centralized proxies or relays, existing systems do not scale with the number of connected clients.
The reality is even worse, not only do centralized proxies not scale, but they literally fall short in front of the request rate limitation strategies adopted by search engines to block bots.
Furthermore, search results are not always accurate as the system may fail in filtering related to fake queries.
Finally, existing systems protect all search queries with a similar and static protection level, by sending the same amount of fake queries for all user queries. 
This strategy is not always effective as user queries may have different levels of sensitivity thus, non-sensitive queries may be overprotected, while sensitive information about the user may be under protected. 
Hence, from our analysis of the state of the art (see Section~\ref{sec:background}), proposing effective private Web search mechanisms requires dealing with essential challenges among which: 
\enumA~enforcing user \textit{privacy} by actually protecting against re-identification attacks through unlinkability and indistinguishability;
\enumB~enforcing \textit{accuracy} by providing the users with similar responses to those they would get while querying the search engine directly; and
\enumC~being \textit{scalable} to millions of users while enforcing service availability in presence of query rate limitation strategies set up by search engines.

In this paper, we present \name{}, the first decentralized private Web search solution that deals with the above challenges as follows.

\mypar{\ding{228} Enforcing privacy} \name{} enforces unlinkability between queries and their originating users as well as indistinguishability between real and fake queries. Specifically, to enforce unlinkability between a query and her sender, each node participating in \name{} acts both as a client when sending own requests and as a proxy by forwarding requests on behalf of other nodes. 
As nodes are controlled by other users, they are regarded as \emph{untrusted}, i.e. query information is not leaked to them.
Therefore, \name utilises trusted execution environments~(TEEs) as provided by Intel SGX~(see Section~\ref{ssec:sgxbackground}). 
Furthermore, it uses secured connections for securing inter-enclave communications and interactions with the search engine.

To mitigate user re-identification attacks, \name sends both fake queries and the real user query through multiple paths to the search engine.
However, instead of blindly sending the same amount of fake queries regardless of the real query, \name is the first private Web search mechanism leveraging query sensitivity. In \name{}, query sensitivity is defined using two dimensions \enumA~\textit{linkability}, which is related to  the similarity of the current request with the user local profile (the higher the similarity the higher the risk of user re-identification); and
\enumB~\textit{semantic sensitivity}, which relates to the topic of the query. Users pick a subset of topics they consider as sensitive of a predefined set. 

Hence, each time a user sends a query to the search engine, \name{} checks whether the query is linkable to her profile and whether the topic of the query belongs to the sensitive topics declared by the her. It then accordingly adjusts the amount of fake queries sent to better protect the request. As such, and contrary to state-of-the-art solutions, \name strongly protects sensitive queries while avoiding to overload the system with fake queries for non-sensitive ones.

\mypar{\ding{228} Providing accurate results} In order to be able to collect accurate responses for the client, \name{} sends fake queries using different forwarding relays than the ones used for forwarding the real query. 
This simplifies the filtering of responses corresponding to fake queries and enables \name{} to return the same responses to the user as when directly querying the search engine, hence reaching a perfect accuracy.

\mypar{\ding{228} Reaching scalability}
In contrast to its closest competitors that rely on centralized proxies,
\name's decentralised architecture consisting of nodes of equal roles leads the system scaling well with growing number of clients.
In practice, we show in Section~\ref{sec:evaluation} that centralized private Web search mechanisms (e.g., \peas~\cite{petit2015peas} and \xsearch~\cite{benmokhtar:hal-01588883}) are not realistic as they get easily blocked by search engines that have aggressive anti-bot strategies. Instead, \name can easily overcome this limitation as the load gets evenly distributed between the participating nodes.

We implemented \name and exhaustively evaluate it on physical machines and using a real workload of query logs extracted from the AOL dataset~\cite{pass2006picture}.
Results show that \name meets expectations. Specifically, we show that: 
\enumA~\name resists re-identification attacks better than its competitors with a low re-identification rate of 4\%; 
\enumB~\name enables sub-second response times, which is on average 13$\times$ faster than using \tor;
\enumC~\name peers can sustain a throughput higher than 40,000~req/s, enabling parallel users to securely browse the search engine; and \enumD~in a complementary simulated deployment involving the 100 most active clients from the AOL dataset, \name fairly balances the load between the participating nodes enabling all users to securely query the search engine without reaching the rate limitation of the search engine.

The remainder of the paper is structured as follows: Section~\ref{sec:background} reports on background and related work while Section~\ref{sec:model} describes our system model. 
Sections~\ref{sec:overview} and~\ref{sec:solution} then present \name. Further, Sections~\ref{sec:SecurityAnalysis}, \ref{sec:experiment} and \ref{sec:evaluation} present the security analysis, our evaluation setup and the results of our experiments.
Finally, Section~\ref{sec:conclusion} concludes this paper and draws some future research directions.

\begin{figure}[!t]
	\centering
		\includegraphics[width=2.3in]{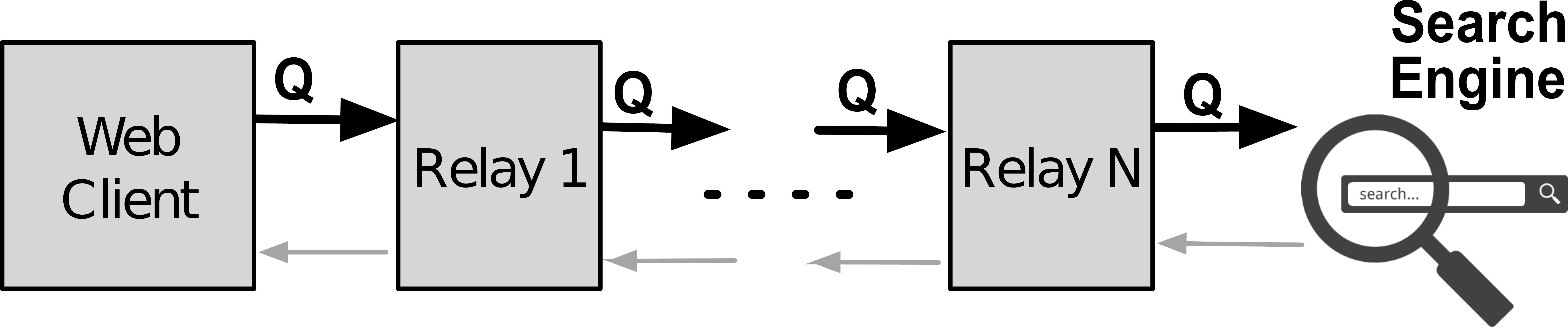}
	\captionof{figure}{Enforcing unlinkability using TOR}
	\label{fig:tor}
\end{figure}

 \section{Related Work and Background}\label{sec:background}
In this section, we first analyse the related work on private Web search solutions (Section~\ref{subsec:privatebackground}), and then give background information on SGX operation principles and system support (Section~\ref{ssec:sgxbackground}).

In the past, several approaches have been proposed to protect data privacy. Homomorphic encryption schemes~\cite{Gentry:2009:FHE:1536414.1536440}
allow computations on encrypted data, without needing access to its plaintext.
In multi-party computations~\cite{du2001secure}, computations are cooperatively conducted among multiple parties while protecting each party's input.
Private Web search solutions that rely on these techniques have previously been proposed in the literature (e.g., \cite{pang2010embellishing}). However, we do not consider these solutions as they require the users to use a novel privacy-preserving search engine (e.g., that would rely on homomorphic encryption to answer queries in a privacy preserving way), while our aim is to build techniques enabling users to benefit from their favourite search engine while preserving their privacy. We review this last category of research work in the following section.

\subsection{Related Work on Private Web Search Solutions}\label{subsec:privatebackground}

Private Web search has been at the heart of active research in the past decade. In this section we analyse existing solutions by starting with the privacy guarantees they offer (Section~\ref{subsubsec:unlink} and ~\ref{subsubsec:indis}) followed by a discussion about their accuracy (Section~\ref{subsubsec:acc}) and scalability (Section~\ref{subsubsec:scal}). We will then conclude with a set of open challenges (Section~\ref{subsubsec:challenges}).  

\subsubsection{Enforcing Unlinkability}\label{subsubsec:unlink}
The first solutions that have been used for privately querying search engines rely on the use anonymous communication protocols to enforce unlinkability between a query and her sender. 
In this context, the most widely used solution is \tor~\cite{dingledine2004tor}, which implements the onion routing protocol~\cite{goldschlag1999onion}. 
As shown Figure~\ref{fig:tor}, each query in \tor is routed through multiple relays using a cryptographic protocol (not shown in the figure). 
\begin{figure*}[tb]
	\begin{subfigure}{0.15\textwidth}
		\centering
		\includegraphics[width=1.2in]{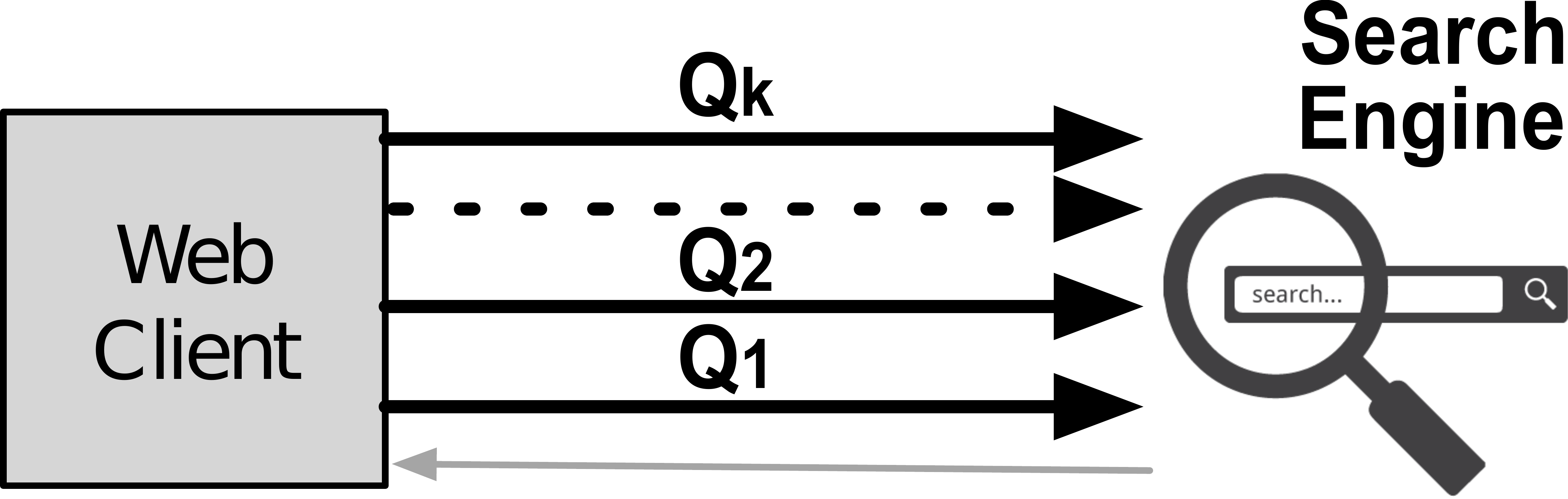}
		\caption{\tmn}
		\label{fig:indist:tmn}
	\end{subfigure}
		\begin{subfigure}{0.24\textwidth}
		\centering
		\includegraphics[width=1.3in]{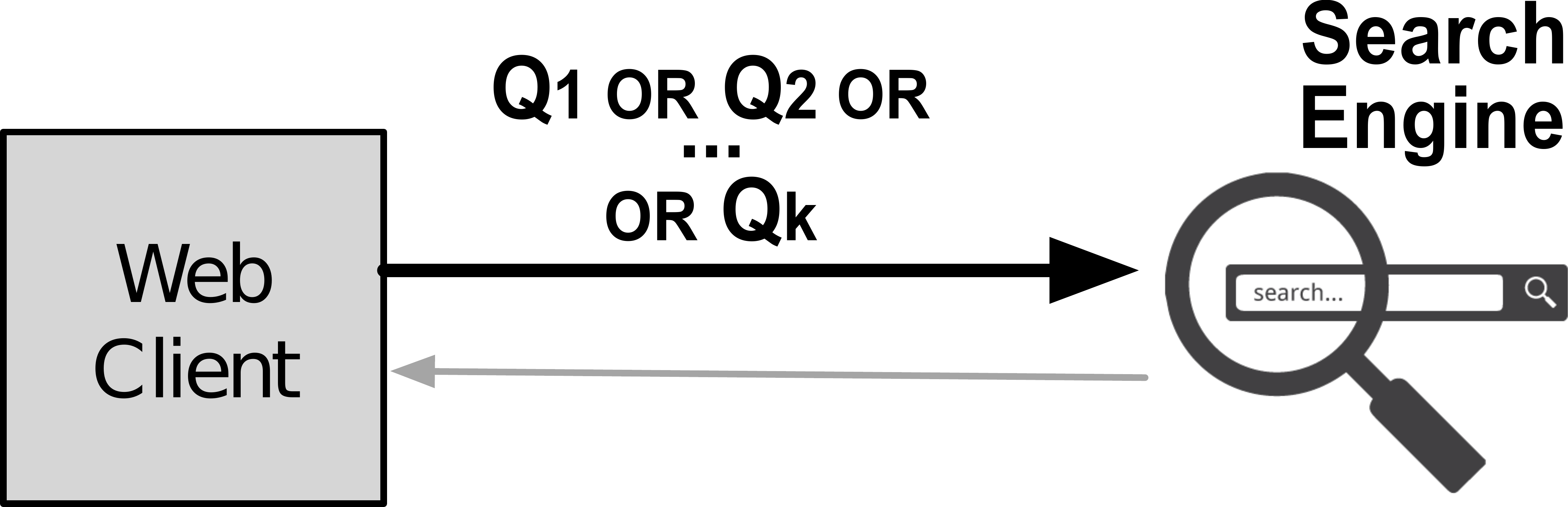}
		\caption{\goopir}
		\label{fig:indist:goopir}
	\end{subfigure}
		\begin{subfigure}{0.22\textwidth}
		\centering
		\includegraphics[width=2.3in]{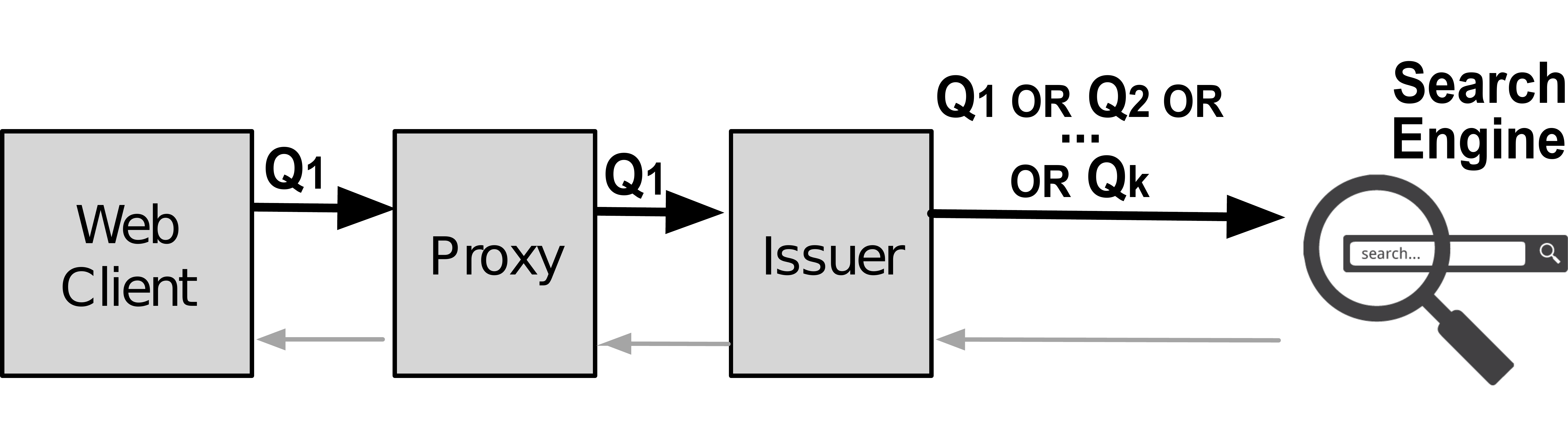}
		\caption{\peas}
		\label{fig:comb:peas}
	\end{subfigure}
		\begin{subfigure}{0.5\textwidth}
		\centering
		\includegraphics[width=1.6in]{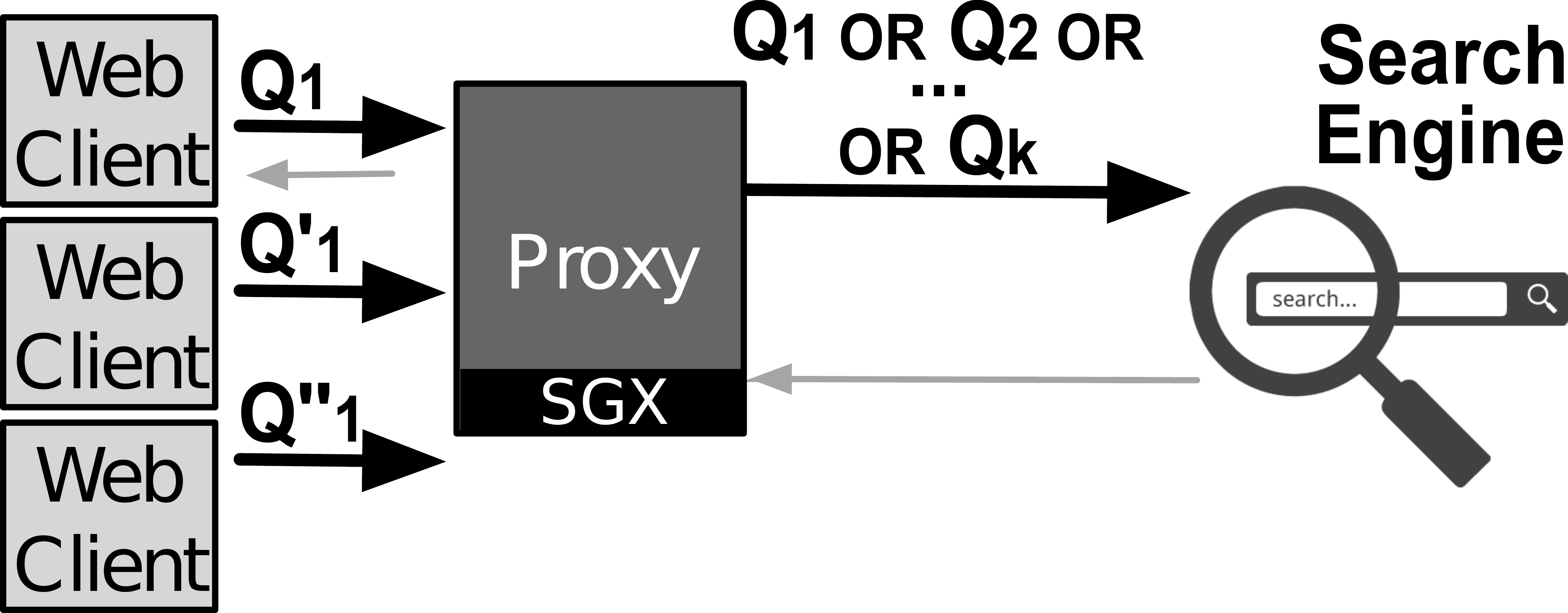}
		\caption{\xsearch}
		\label{fig:comb:xsearch}
	\end{subfigure}
	\caption{Systems enforcing indistinguishability (a,b) and combining unlinkability and indistinguishability (c,d)}
	\label{fig:indist}
	\vspace{-0.5cm}
\end{figure*}

Specifically, each query is encrypted using the public keys of a set of nodes randomly selected to act as relays creating an \emph{onion} with multiple encryption layers. This onion is then successively routed through the selected relays. Upon receiving the onion, each relay deciphers its outer layer using its private key and forwards the inner onion to the following relay, until the onion reaches the exit node. The exit node retrieves the query and sends it to the search engine on behalf of the user. Other protocols increasing the security of \tor (e.g., against relays acting selfishly) have been proposed in the literature (e.g., Dissent~\cite{wolinsky2012dissent}, RAC\cite{mokhtar2013rac}) but we do not discuss these alternatives as their cost (mainly due to the use of all-to-all communication primitives and heavy cryptography) makes them impractical for private Web search.

Most importantly, all these protocols, including \tor (see Section~\ref{sec:evaluation}) are not resilient to re-identification attacks~\cite{peddinti2014web,petit2016simattack}. These attacks work as follows: assuming a set of user profiles built from user past queries, user re-identification attacks try to link anonymous queries to a profile corresponding to their originating user. These attacks are successful even when the users rely on anonymous communication protocols because users tend to look for similar things even when they have a different identity (i.e., IP address) on the Internet.
Additionally, as we show in Section~\ref{sec:evaluation}, \tor induces a high end-to-end latency of several seconds, which heavily impacts user experience.

\subsubsection{Enforcing Indistinguishability}\label{subsubsec:indis}

Alternative private Web solutions have been proposed (e.g., TrackMeNot~\cite{howe2009trackmenot}, GooPIR~\cite{domingo2009h}). These solutions depicted in Figure~\ref{fig:indist} aim at making real user interests indistinguishable from fake ones. Specifically, TrackMeNot (Figure~\ref{fig:indist:tmn}) is a browser extension, which periodically sends fake queries to the search engine on behalf of the user. Hence, eventually, the user profile stored at the search engine gets obfuscated mixing the user real interests with fake ones. Instead, GooPIR (Figure~\ref{fig:indist:goopir}) obfuscates each user query by aggregating $k-1$ fake queries with the real one using the logical $OR$ operator. As such, the search engine can not distinguish the real query from fake ones. However, these solutions suffer from two limitations:
\enumA~the identity of the user is known to the search engine; and 
\enumB~they have been subject to attacks (see Section~\ref{sec:evaluation}) as the fake queries they generate (based on RSS feeds and dictionaries) are easily distinguishable from real ones.

To overcome these limitations two recent solutions combining unlinkability and indistinguishability have been proposed in the literature. The first one called \peas~\cite{petit2015peas} (Figure~\ref{fig:comb:peas}) is based on two non-colluding servers. The first server, called the \emph{proxy}, has access to the identity of the requester but has not access to the content of the query as the latter is encrypted with the public key of the second server. Instead, the second server, called the \emph{issuer}, has access to the query but does not know the originating user. In addition to forwarding the query on behalf of the user, the issuer generates $k-1$ fake queries and aggregates them with the original query to enforce indistinguishability. Differently from \goopir and \tmn, \peas's fake queries are generated using a co-occurrence matrix of terms built by the issuer from other users past queries. Hence, \peas better resists re-identification attacks as its fake queries are syntactically closer to real ones. 

More recently, a novel solution to private Web search called \xsearch (Figure~\ref{fig:comb:xsearch}), improving \peas has been proposed.  \xsearch enforces both unlinkability and indistinguishability and
builds upon Intel SGX enclaves to run query obfuscation on untrusted proxy nodes.

One of the limitations of the above solutions is that all user queries get obfuscated with the same intensity (e.g., by generating $k-1$ fake queries in \goopir, \peas and \xsearch) regardless of their sensitivity. 
Indeed, a small value of $k$ may lead to under protecting sensitive queries, which increases the risk that they get linked back to the original user while a large value of $k$ may unnecessarily generate a large amount of traffic. 

\subsubsection{Accuracy of Query Results}\label{subsubsec:acc}
Enforcing privacy always comes at a cost. In the context of private Web search, enforcing indistinguishability by aggregating the user query with fake queries (e.g., using the $OR$ operator) generates noise in the responses sent by the search engine as the responses corresponding to fake queries get merged with those corresponding to the real one. This noise is generally filtered out at the client side (in \peas and \goopir) or by the proxy (for \xsearch) by removing the responses that do not contain words composing the original query. However, as further quantified in Section~\ref{sec:evaluation}, despite this filtering process, relevant responses of the original query may be lost, while noise coming from fake queries may be returned to the user. 
Furthermore, the logical $OR$ operator for multiword-based queries is not natively supported by all search engines and is impractical as the search engine returns results only related to the exact query,  
with a direct impact on the accuracy of the corresponding private Web search mechanism. 
\subsubsection{Scalability}\label{subsubsec:scal}
Web search is with no doubt one of the most used service over the Internet. Hence, it is not possible to aim for a private Web search mechanism to be used in practice if it does not scale to millions of users. This is not the case of centralized mechanisms such as \peas or \xsearch even though their authors discuss the possibility to move to distributed deployments. In addition to the ability of private Web search to sustain the load coming from Internet users, a more concrete problem comes from the rate limitations imposed by search engines to counter bots and other attacks. 
For instance, our experience with querying Google from the used prototype shows that after a high flow of queries, Google's bot protection triggers and asks to fill a captcha.
Another problem of approaches like \peas and \xsearch is the deployment of proxies that generate costs. In contrast, \name leverages client machines,
thus, no deployment is necessary.
These concrete limitations motivate the distributed design of \name as further discussed in Section~\ref{sec:overview}.

\subsubsection{Summary of Open Challenges for Private Web Search}\label{subsubsec:challenges}
From the analysis of state of the art private Web search solutions (also summarized in Table~\ref{tab:summary}), it appears that there is no solution for enforcing privacy (both unlinkability and indistinguishability) and scalability while providing accurate responses to the users. Before describing how \name{} addresses these challenges (in Section~\ref{sec:overview}), we first introduce  preliminaries about Intel SGX enclaves that we leverage in this work for preventing information leakage on untrusted nodes.

\begin{table}[t]
    \setlength{\tabcolsep}{2.3pt}
    \footnotesize
\centering
   \begin{tabular}{r c  c  c  c  c  c }
   \toprule
	&\tor	&	\tmns &	\goopir  &	\peas &	\xsearch	& 	\SYS	\\
	\midrule
Unlinkability & \y	&		\n		&		\n		&		\y		&	 \y		&  \y\\
Indistinguishability & \n	&		\y		&		\y		&	\y			&	 	\y	& \y\\
Accuracy & \y	&		\y		&		\n		&		\n		&	 \n		& \y\\
Scalability & \y	&			\y	&		\y		&		\n		&	\n 		& \y\\
	\bottomrule
   \end{tabular}
   
	\caption{Comparison of private Web search mechanisms.}	
	\label{tab:summary}
\end{table}

\subsection{Background on Intel SGX and System Support}
\label{ssec:sgxbackground}

\SYS uses Intel SGX to establish trust in usually untrusted nodes.
SGX provides trusted execution environments~(TEEs) called \emph{enclaves} firstly introduced by Intel with the Skylake architecture.
Applications create such enclaves to protect the integrity and the confidentiality of the data and the code being executed.

Memory pages associated with enclaves are stored in the \emph{enclave page cache}~(EPC) and are integrity-protected and encrypted by the \emph{memory encryption engine} (MEE)~\cite{gueron2016memory}.
Thus, SGX withstands even physical attacks on enclave memory: a memory dump will always produce encrypted data.
However, the EPC is limited to $128$~MB due to hardware restrictions.
If this limit is exceeding, enclave pages are subject to a swapping mechanism implemented in the Intel SGX driver, resulting in a severe performance penalty~\cite{brenner2016securekeeper,arnautov2016scone}.
Note that future releases of SGX might relax this limitation~\cite{mckeen2016intel}.
Furthermore, Intel SGX provides \emph{remote attestation}, allowing a remote third party to verify that an enclave runs on a genuine Intel processor with SGX. After successful remote attestation, secrets such as keys can securely be injected into the enclave.

\begin{figure}[t]
\vspace{-0.3cm}
  \centering
  \includegraphics[scale=1.0]{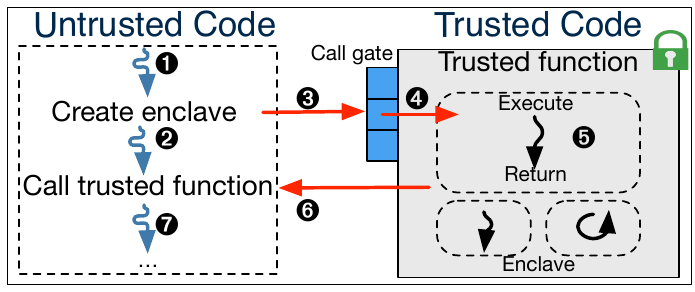}
  \caption{SGX operating principles.}
  \label{fig:sgx}
\end{figure}

Intel provides a software development kit~(SDK)~\cite{intel-sdk} to support developers writing SGX applications.
The SDK can manage the life cycle of enclaves and introduces the notion of \texttt{ecalls} (calls into enclaves) 
and \texttt{ocalls} (calls out of enclaves).
Figure~\ref{fig:sgx} depicts the basic execution flow of SGX.
First, an enclave is created \ding{202}.
As the program must execute a trusted function \ding{203}, it performs an \texttt{ecall} \ding{204},
passing the SGX call gate to bring the execution flow inside the enclave.
The trusted function is then executed by one of the enclave's threads \ding{205}.
The result is encrypted and returned \ding{206} to give back the control to the untrusted main thread.

Current web browsers lack system support for trusted execution of the one offered by SGX.
However, this support can be retrofitted using browser extensions that have the ability to call native code, thus also invoke calls into and handle ocalls from SGX enclaves.
The authors of TrustJS~\cite{goltzsche2017trustjs} follow this approach to enable client-side trusted execution of JavaScript.
They modify a JavaScript interpreter to run inside an SGX enclave and integrate it using a browser extension into the Firefox browser.
\name follows a similar direction and integrates SGX enclaves into commodity browsers.
Instead of protecting JavaScript execution from potentially malicious users, it targets privacy preserving decentralized web search as a service carried out by users for users.   

\section{System and Adversary Model}
\label{sec:model}

Before presenting the design principles of \name{} in Section~\ref{sec:overview}, we describe our assumptions and the adversary model against which our protocol is designed.

First, we assume that each \name node supports the use of Intel SGX instructions. Given the plan of Intel to include the SGX technology in all major future Core CPU releases~\cite{intel-8th-gen}, we believe this is a reasonable assumption. However, we assume that SGX behaves correctly, i.e., there are no bugs or backdoors.
Additionally, we do not deal with side-channel attacks against SGX \cite{weichbrodt2016asyncshock,xu2015controlled}.
We consider such attacks as outside the scope of this paper and that the research community provides solutions \cite{shih2017t,fu2017sgx}
that might eventually be incorporated in SGX.
Furthermore, we are also aware that denial-of-service~(DoS) attacks cannot be prevented, e.g. malicious clients might not initialise the enclave, invoke calls into enclaves or drop all queries.
Finally, we assume that all the used cryptographic primitives and libraries are trusted and can not be forged.

The computations performed by \name{} for each user query go through three premises, namely:
\enumA~the client machine;
\enumB~a set of remote peers; and
\enumC~the search engine.
These are subject to different levels of trust:

First, we assume that the client machine issuing search requests is trusted.
This includes all the computations performed locally outside of enclaves and that involve client information (e.g., locally assessing the sensitivity degree of a query). 
Furthermore, we assume that the local communication between a human user and \name{} is trusted: that is an adversary can not modify the query that the human user has typed, nor it can modify the local configuration of \name{} that the human user has set up (i.e., the set of topics she considers as semantically sensitive). 
Note that it is possible to relax or even remove this assumption by integrating the work from~\cite{Weiser:2017:SGT:3029806.3029822} for trusted I/O. 

Second, however, users do not trust the remote peers used as relays to forward their queries. 
Specifically, we assume that remote peers can act in a Byzantine manner \cite{pease1980reaching,lamport1982byzantine}: they can behave arbitrarily by crashing, being subject to bugs or being under the control of malicious adversaries.

Third, we assume that the search engine is honest but curious. 
That means, it faithfully replies to search queries while gathering information from incoming queries, but is able to build user profiles and run re-identification attacks~\cite{petit2016simattack}.
 
\begin{figure}[t]
\centering
\includegraphics[scale=1.0]{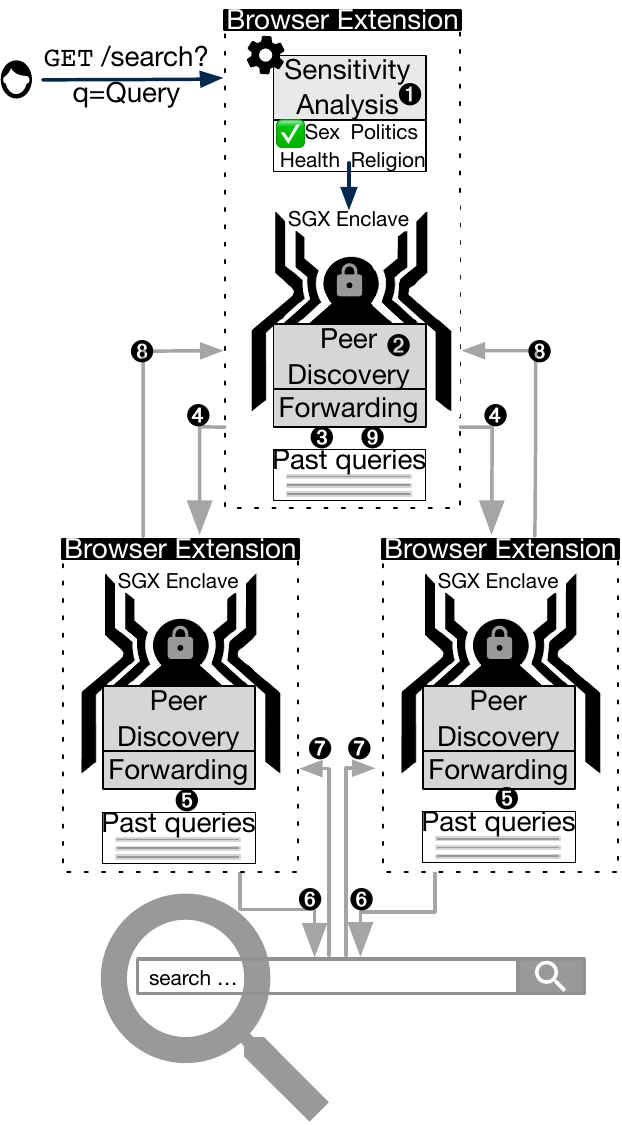}
\vspace{-2mm}
\caption{\SYS architecture and operating flow.}
\label{fig:picture}
\end{figure}

\section{\name{} in a Nutshell}
\label{sec:overview}

To efficiently protect users during Web search, \SYS combines both \textit{unlinkability} and \textit{indistinguishability}, two complementary properties described in Section~\ref{sec:background}.
The former hides the identity of the requesting user by routing her queries to the search engine through other nodes in the system. The latter makes the real query indistinguishable among other fake queries. We briefly introduce in this section how these two properties are enforced in \SYS before giving details in Section~\ref{sec:solution}.
 
To use \SYS, a user has to install the \SYS browser extension. As such, users seamlessly get protected without changing their browsing habits, i.e., using a Web browser. Then, each time the user formulates a query $Q_u$, the latter is processed as follows (and also depicted in Figure~\ref{fig:picture}).

First, the \SYS browser extension evaluates the sensitivity of the query (step \ding{182} in the figure).
To do so, \SYS follows a user-driven approach, and combines a semantic-based approach and a linkability analysis (Section~\ref{ssec:sensitivity}). 
The sensitivity analysis produces a score $k$, that is the number of fake queries used to make the real user query $Q_u$ indistinguishable from others queries. 

Then, by relying on a peer discovery component (step \ding{183} in the figure also discussed in Section~\ref{ssec:peerdiscovery}), \SYS selects $k+1$ random peers $P_{p0},P_{p1},...,P_{pk}$ to which it sends $k$ fake queries $Q_{p1},...,Q_{pk}$ and the real query (step \ding{184} in the figure). In \SYS, fake queries are generated from past queries issued by other users in the system and that are stored in a local table each time a node acts as a relay for other nodes. Using past user queries as fake queries makes them look more real than those generated by systems such as \tmn or \goopir where fake queries are generated using RSS feeds or dictionaries. 

Then, when a request is received by a peer acting as a relay, the latter is stored in the local table of past queries (step \ding{186}) before it is sent to the search engine (step \ding{187}). In \SYS, real queries and fake ones are processed similarly by the relays. Hence, an external observer analysing the (encrypted) network traffic has no clue whether a node is sending out a real query, a fake one or whether he is forwarding someone else's query, which is not the case of systems where fake queries are generated at the relays (e.g., \xsearch or \peas). In these systems, even though the traffic is encrypted, an adversary can infer whether an outgoing message is a real query or an obfuscated one from the request size (e.g., messages containing obfuscated queries using the $OR$ operator are larger than messages containing the real query). Query forwarding is further described in Section~\ref{ssec:QueryForwarding}.

Upon receiving a request from the node acting as a relay, the search engine sends the answers to the latter node (step \ding{188}), which routes the responses to the initial sender \ding{189}.
Finally, the \SYS forwarding component drops the responses corresponding to fake queries (step \ding{190}) before the browser extension displays the result of the real query to the user.	

To avoid information leakage, all components of \name that process sensitive data (in our context queries issued by other users) are located within the enclave. Instead, in order to minimize the amount of trusted code, components that process data related to the user who owns the machine are run outside the enclave. For instance, the part in charge of assessing the sensitivity of queries issued by the local user is performed outside the enclave as we trust the client machine where the search queries are issued (see Section~\ref{sec:model}). This allows to drastically minimise the amount of trusted code, which reduces the risk of having of critical bugs.
However, as we do not trust remote peers, parts that handle the queries of other users in plain text are located inside the enclave.
Furthermore, all messages being exchanged by these parts between different \name nodes are encrypted and decrypted within the enclave.
More precisely, peer discovery, query forwarding as well as en- and decryption are executed within the enclave.
Finally, as \name stores past user queries to generate fake queries, these are stored in enclave memory.

\section{Detailed Description of \name}
\label{sec:solution}

In this section, we first present the sensitivity assessment performed in \SYS (Section~\ref{ssec:sensitivity}). We then present the dynamic query protection and the forwarding scheme (Sections~\ref{ssec:AdaptiveQueryProtection} and \ref{ssec:QueryForwarding}). Finally, we present how a client bootstraps \SYS and the peer discovery protocol before presenting  implementation details Section~\ref{ssec:implementation}).

\subsection{Sensitivity Analysis in \SYS}
\label{ssec:sensitivity}

To improve indistinguishability of queries while not overloading the network at the same time, 
\name dynamically protects user queries according to their actual sensitivity.
To measure the sensitivity of a query, \name relies on two risk assessment measures computed outside the enclave: 
\enumA~the semantic assessment; and
\enumB~the linkability assessment.
As further discussed in Sections\ref{sssec:LexicalSemanticsBasedAnalysis} and \ref{sssec:LinkabilityAnalysis}, the former analyses the actual topic of the query with respect to sensitive topics declared by the user while the latter assesses the risk that the query gets linked back to its originating user by comparing it with other queries previously sent by the user. 

\subsubsection{Semantic-based Analysis}
\label{sssec:LexicalSemanticsBasedAnalysis}

The semantic-based assessment aims at identifying semantically-sensitive queries. 
As semantic sensitivity is subjective (one query might be considered as sensitive by one user and non-sensitive by another user), \name proposes a user-centric approach where each user selects a set of topics that she considers as sensitive.
To define sensitive topics, we use the privacy policy of Google which defines sensitive personal information as \textit{"confidential medical facts, racial or ethnic origins, political or religious beliefs or sexuality"}~\cite{google-terms}.
Consequently, by default a user in \SYS can select sensitive categories among health, politics, sex, and religion.
Nevertheless, a user can import dictionaries to create other sensitive topics as described below.

The semantic-based assessment is binary and defines if the query belongs to at least one topic marked as sensitive.
To achieve that, we use two complementary information retrieval approaches to build a dictionary of terms associated to each identified sensitive topics.
The first approach uses two libraries: 
\enumA~WordNet, a lexical database, and
\enumB~eXtended WordNet Domains, a mapping of WordNet synsets to domain labels. 
A synset is a set of synonym words. 
We use the categories defined in the eXtended WordNet Domains library which are related to our privacy-sensitive topics.	
We then use the mapping of these categories with WordNet synsets to identify all keywords related to each sensitive topic.
The dictionaries built for each sensitive topic gather these keywords. 
The second approach captures statistical correlations among words and sensitive topics represented using latent topic models (i.e., Latent Dirichlet Allocation, LDA~\cite{Blei:2003:LDA:944919.944937}).
This generative probabilistic model is well adapted for modelling text corpora. 
In the LDA model, a sensitive topic is described through different thematic vectors that indicate the latent dimensions of the topic.
Once this model is trained with a text corpora associated to the considered sensitive topics, we build the dictionary by gathering all terms of all thematic vectors.
Section~\ref{ssec:implementation} gives implementation details for the semantic-based assessment based on WordNet and LDA.

\subsubsection{Linkability Analysis}
\label{sssec:LinkabilityAnalysis}
The goal of the linkability assessment is to determine if the
query is vulnerable to a re-identification attack. In such attacks,
an adversary tries to link an anonymous query to a specific
user by measuring the distance between the query and a set of
user profiles built from past user queries (e.g., collected when
the users were not using private Web search mechanisms).
Concretely, the linkability assessment, which is performed
on the client side, provides a score in $[0,1]$ measuring the
proximity of the current query to past user queries already
sent to the search engine. To do that, we first represent the
query q in a binary vector where each element of the vector is
a term in the query. We then compute the cosine similarity between the vector associated to the query and the vector of each past queries issued by the user. Finally, to give more importance to past requests that are similar to the current user request compared to non-similar ones, the results of the cosine similarities are ordered and an aggregated value is computed using exponential smoothing.

\subsection{Adaptive Query Protection}
\label{ssec:AdaptiveQueryProtection}
\name leverages the sensitivity analysis to dynamically adapt the query protection. 
More precisely, both the semantic-based and the linkability assessments control the obfuscation scheme of \name, the more sensitive a query is, the more protected it will get.
Specifically, if the query includes at least one term which belongs to a dictionary related to a sensitive topic defined by the user, 
the number of fake queries is maximal, as defined by $k_{max}$. This behaviour minimizes the risk of re-identification for queries related to sensitive topics.

For queries that are not semantically sensitive, the number of fake queries (value of $k$) is defined according to a linear projection between the score returned by the linkability assessment in $[0,1]$ and the maximum number of fake queries, $k_{max}$.
This scheme dynamically adapts the protection according to the actual risk of re-identification attack.

\subsection{Query Forwarding}
\label{ssec:QueryForwarding}
\vspace{-2pt}

Once the number of fake queries (noted $k$) is decided according to the actual sensitivity of the user query, the process continues in the SGX enclave as it involves remote peers.
More precisely, \name makes the real query indistinguishable by choosing fake ones and unlinkable by routing them to the search engine through different paths.
\name uses peer discovery to dynamically maintain a random view of other alive nodes in the system running \name as further described in Section~\ref{ssec:peerdiscovery}.
Then, \name picks random $k+1$ nodes of this random view to act as proxies for the $k$ fake queries and the original one.
By using this random view, we additionally ensure a load balancing over all nodes in the system.

Then for each of the $k+1$ random nodes, a query is randomly selected in the table of past queries excepted once where the original query is selected.
Then each of their $k+1$ queries are respectively forwarded to their $k+1$ selected proxies.
The identity of the proxy dealing with the original query is maintained in a table.
All forwarded queries are encrypted before being send to other peers.

Once a proxy receives a query forwarding request, it adds this query in its local table of past queries and routes this query to the search engine.
The received answers from the search engine are returned to the original user.
Finally, on the original client, the answers received from the proxy that managed its original query are presented to the user.
The answers received from other proxies are silently dropped.

\subsection{Bootstrapping \SYS}
Besides declaring its sensitive topics, there are three key elements that need to be bootstrapped when \SYS is first launched by a given user. First, when the system is first started, there are no past queries stored in the enclave to be used as fake queries. Hence, \SYS fills the fake queries table using popular Google queries~\cite{google-trends}. Queries extracted from this Web site reasonably look as real queries as they are issued by real users regarding trendy topics. Then, the \SYS browser extension has to bootstrap peer discovery. We assume that bootstrapping the peer discovery protocol is done as in classical peer-2-peer systems using a public repository of IP addresses (e.g., as in \tor) from which a \SYS instance can select a first sample of random peers. This sample with then be periodically shuffled using the peer discovery protocol (Section~\ref{ssec:peerdiscovery}).  
Finally, the remote attestation mechanism need to be initialized. Remote attestation facilities provided by Intel SGX are used by \SYS to authenticate remote enclaves to each other. This ensures that
\enumA~nodes only talk to genuine Intel SGX enclave; and
\enumB~all enclaves are known implementations.
To do so, while bootstrapping, a \name client challenges every connecting enclave to send a so-called \emph{quote}.
This data structure contains the a hash of the enclave code and a secret for applying encryption.
The quote is checked for a known hash value and is transferred to the trusted Intel attestation service~(IAS) for verification if it originates from a genuine SGX platform.

\subsection{Peer Discovery in \SYS}\label{ssec:peerdiscovery}
Peer discovery in \SYS is done using classical algorithms and contributions to this field fall outside the scope of the paper. Specifically, the selection and maintanance of random views is using the random-peer-sampling protocol~\cite{Jelasity:2007:GPS:1275517.1275520} which ensures connectivity between nodes by building and maintaining a continuously changing random topology.

\subsection{Implementation Details}
\label{ssec:implementation}

To allow end users to integrate \name seamlessly into their workflow, we designed it as an extension to the Firefox browser.
The JavaScript-based extension integrates the \name SGX enclave using \emph{js-ctypes},
allowing asynchronous calls to and from the enclave into the untrusted extension code.

\name uses TLS connections to search engines.
These connections need to be established from within enclaves in order to not disclose queries of other user to untrusted machines.
\name implements this by linking the enclave code to an SGX-compatible version of mbedTLS~\cite{mbed-tls}.
Adding this library results in an enclave of only $1.7$~MB, thus, \name does not suffer from EPC paging.

To automatically detect semantically sensitive queries, \name relies on both WordNet libraries and LDA statistical modelling. 
WordNet's machine-readable lexical database is organized by meanings, where words are grouped into sets of synonyms called synsets~\cite{fellbaum98wordnet}.
The eXtended WordNet Domains library maps every WordNet synset to 170 domain labels.
For the experiments described in this paper, we consider sexuality as an example of sensitive subject in user queries.
We trained a LDA statistical model using the Mallet toolkit~\cite{McCallumMALLET}, with 200 topics on 2M of titles and descriptions of videos related to the sensitive subject~\cite{mazieres:hal-00937745}.
Finally, every query including a term present in at least one LDA topic or linked to WordNet domain related to sensitive subject is identified as semantically sensitive.

\section{Security Analysis}
\label{sec:SecurityAnalysis}
This section presents our security analysis of \SYS. 
We consider threats, either from the point of view of a client, a \SYS proxy, or from the web search engine perspective.

\paragraph{On the client side}
Clients can not bypass the SGX enclave.
Indeed, \name relies on keys being generated during start-up in the enclave. The keys are only exchanged with
other genuine SGX enclaves after successful remote attestation. Therefore, clients that attempt to 
bypass enclaves cannot create correctly encrypted/signed requests.
The threat model of \name (Section~\ref{sec:model}) assumes the client node as trusted.
If the client is compromised, the sensitivity analysis could be subverted.
However, the choice of the proxies and the forwarding process as well as the table of past queries can not be subverted as they are inside the SGX enclave. 

\paragraph{On the proxy side}
Inter-enclave traffic as well as the traffic between the enclave and the search engine are protected through encrypted channels.
In addition, all forwarding performed by the proxy is done inside the SGX enclave.
Consequently, a malicious proxy cannot hamper \name, as we do not consider side-channel attacks as described in Section~\ref{sec:model}.
However, a malicious process could replay user past queries on the proxy. 
This threat can be limited by including a random identifier in each message to detect a replay. 
Also, a malicious proxy can deny initialization or calls into enclaves.
\name solves this by letting clients blacklisting peers that do not respond within a given period of time.

\paragraph{On the search engine side}
As shown in Section~\ref{ssec:EvalPrivacy}, the capability of the search engine to re-identify users is very limited.
However, as fake queries are in fact real past ones, the search engine could identify a real query when it receives this query for the first time. 
But in this case, the identity of the requesting user is still hidden by the proxy. 

\section{Experimental Setup}
\label{sec:experiment}

This section first presents the experimental setup used to evaluate \SYS, which includes competitors, datasets and metrics.

\subsection{Comparison Baselines}
\label{sec:baselines}

We compare \SYS against five state-of-the-art private Web search approaches (\emph{e.g} \tor~\cite{dingledine2004tor}, \tmn~\cite{howe2009trackmenot}, \goopir~\cite{domingo2009h}, \peas~\cite{petit2015peas}, \xsearch~\cite{benmokhtar:hal-01588883}), and a protection-free Web search scenario.
These approaches are further described in Section~\ref{sec:background}.
\subsection{Data Sources}
\label{ssec:Dataset}

We use a dataset of real queries from the AOL query log dataset~\cite{pass2006picture}.
This dataset contains approximately 21 million queries formulated by 650,000 users over a three month period.
We use the same methodology as described in~\cite{Gervais:2014:QWP:2660267.2660367} to evaluate Web-search privacy by considering a subset of the most active users.
These users are the ones that exposed the most information through their past queries, which makes them also the most difficult to protect.
Here, we manually extracted $198$ users among the subset of users who sent at least one semantically sensitive query.

As described in Section~\ref{ssec:MetricPrivacyWebSearch}, an adversary needs a prior knowledge about each user to perform a re-identification attack. 
We split queries into two sets: a training set that represents prior knowledge held by the adversary about the users (2/3 of the dataset), and a testing set that represents new user queries that are protected (the remaining 1/3 of the dataset).
On average this represents \emph{i.e.}~$487.6$ queries per user for the training set over a total of $96,547$ queries. 

\subsection{Crowd-Sourcing Campaign for Query Sensitivity}
\label{sec:crowdsourcing}

To determine user-perceived sensitivity with regard to Web queries, we conducted a crowd-sourcing campaign using Crowdflower~\cite{crowdflower}.
We selected the first $10,000$ queries over all user queries in the testing set (c.f.,~Section~\ref{ssec:Dataset}), and asked the crowd-sourcing workers to determine if these queries are related to sensitive topics.
We considered different sensitive topics (\emph{i.e.},~health, politics, religion, sexuality, others), and each query was annotated by $5$ different workers to obtain multiple opinions.
The result of the campaign is that only $15.74\%$ of the queries are related to sensitive topics.
This motivates the adaptive approach followed by \SYS that applies a dynamic protection scheme to sensitive queries.
\subsection{Measuring Accuracy of \SYS's Query Categorizer}
\label{ssec:MetricAccuracyQueryCategorizer}

To measure the accuracy of \SYS's query categorizer that automatically determines if a query belongs to a sensitive topic, we consider the precision and the recall metrics.
The \textit{recall} calculates the proportion of queries detected as sensitive by \SYS among all actually sensitive queries.
And the \textit{precision} calculates the proportion of actual sensitive queries among queries detected as sensitive by \SYS.
Let $Q$ be the set of queries, $Q_{s}$ the set of actually sensitive queries (i.e., related to sensitive topics), and $Q_{m}$ the set of queries that are identified as sensitive by \SYS's query categorizer.
Recall and precision are respectively defined as follows:

\begin{center} 
\footnotesize
\vspace{-2mm}
$$Recall\textbf{ } = \frac{|Q_{m} \cap Q_{s}|}{|Q_{s}|},  \quad
Precision\textbf{ } = \frac{|Q_{m} \cap Q_{s}|}{|Q_{m}|} $$
\vspace{-4mm}
\end{center}

\begin{table*}
\vspace{-0.3cm}
\begin{minipage}{\linewidth}
      \centering
\hspace{-4mm}
      \begin{minipage}{0.37\linewidth}
\begin{figure}[H]
	\centering
	\includegraphics[scale=0.5]{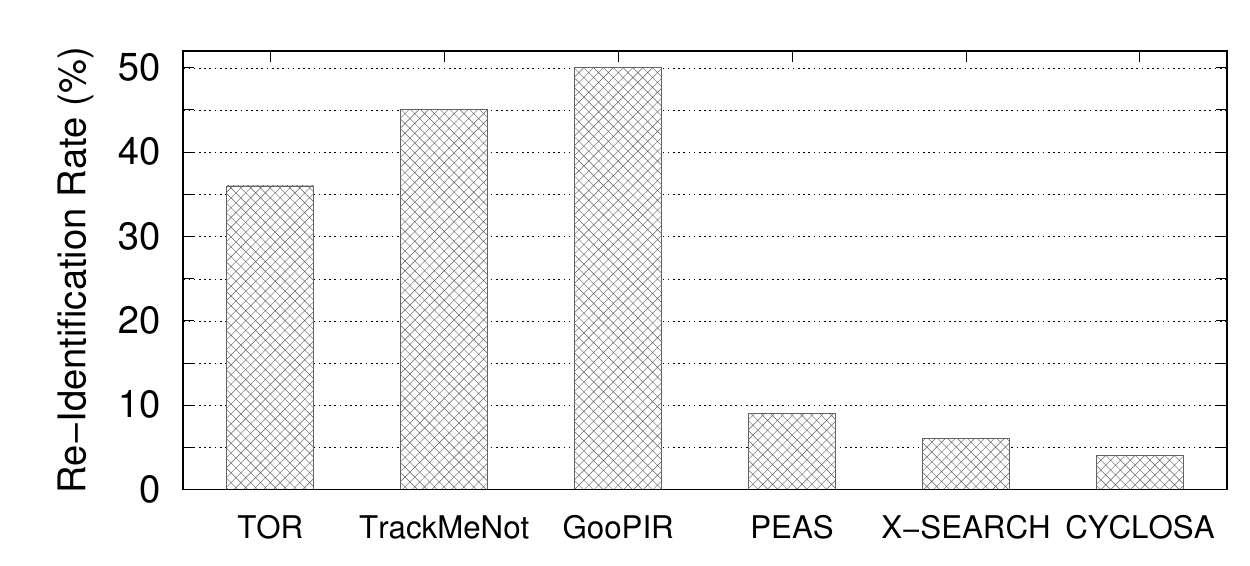}
	\caption{Comparison of \SYS's privacy level with competitors -- The lower re-identification rate, the better privacy.}
	\label{fig:PrivacyEval-ReidentifAttack}
\end{figure}
      \end{minipage}
\hspace{1mm}
	\begin{minipage}{0.37\linewidth}
\begin{figure}[H]
\vspace{-4mm}
	\centering
	\includegraphics[scale=0.5]{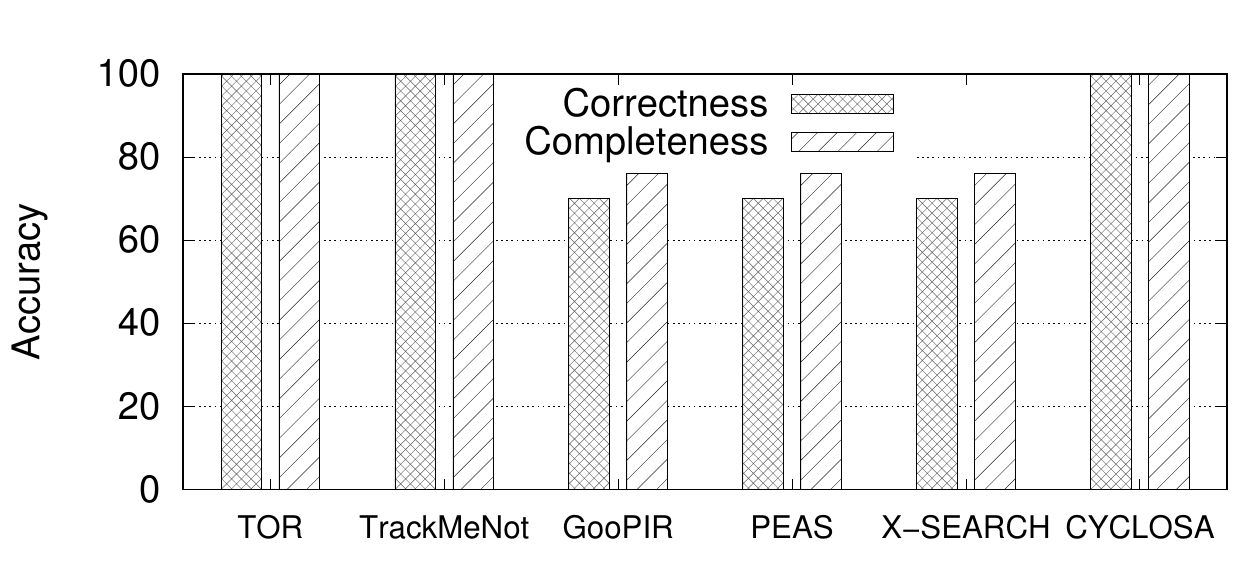}
	\caption{Accuracy of results returned to users for \SYS and state-of-the-art competitors.}
	\label{fig:AccuracyEval-WebSearch}
\end{figure}
      \end{minipage}
\hspace{1mm}
	\begin{minipage}{0.22\linewidth}
\begin{figure}[H]
\centering
\includegraphics[scale=0.55,trim=0 0 0 0]{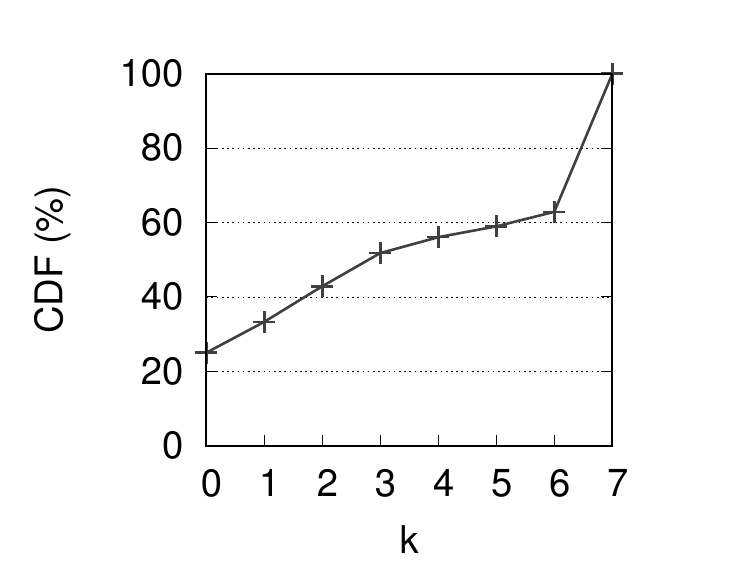}
\caption{Actual number of fake queries in \name.}
\label{fig:kvalues}
\end{figure}
      \end{minipage}
  \end{minipage}
\vspace{-4mm}
\end{table*}

\subsection{Measuring Privacy of Web Search}
\label{ssec:MetricPrivacyWebSearch}

To evaluate privacy, we use the SimAttack user re-identification attack to measure the robustness of privacy preservation solutions~\cite{petit2016simattack}.
Here, we assume an adversary that intercepts queries arriving to the search engine, and that has prior knowledge about each user in the form of a user profile containing user's past queries (i.e.,~from the training set of the dataset presented in Section~\ref{ssec:Dataset}).
Then, based on the sensitive topics chosen by the users and the linkability, the queries of the testing set are adaptively protected by \SYS, before sending them to the search engine through different paths.

SimAttack measures the similarity between a query $q$ and a user profile $P_u$, where $P_u$ contains queries that belong to the training set of user $u$, and the additional knowledge of the attacker when intercepting queries.
This metric accounts the cosine similarity of $q$ and all queries part of the user profile $P_u$, and returns the exponential smoothing of all these similarities ranked in ascending order.
Given a query $q$, the metric is calculated for all users' profiles.
If the metric is higher than $0.5$ to ensure a certain confidence, and if only one user profile has the highest similarities, SimAttack returns the association between that user profile and the query $q$. 
If that user profile actually corresponds to the profile of the user that issued the query $q$, re-identification is successful.
Otherwise, SimAttack responds that re-identification is unsuccessful. 

Thus, to evaluate the level of privacy provided by a Web search privacy protection solution, we use SimAttack to calculate the overall re-identification success rate, 
that is the proportion of queries for which the user profile is successfully re-identified to all queries sent to the Web search when running queries from the testing set (c.f.,~Section~\ref{ssec:Dataset}).
Obviously, the lower re-identification success rate, the better privacy level.

\subsection{Measuring Accuracy of Private Web Search}
\label{ssec:MetricAccuracyWebSearch}

By design, \name returns to users results associated to its search query.
However, as described in Section~\ref{sec:baselines}, other protection mechanisms such as \xsearch and \peas include an obfuscation scheme that impacts the results returned by a search engine.
Consequently, we evaluate the capacity of considered candidates to return results to the user only related to its initial query.
To achieve that, for a given initial query, we compare results returned by the search engine for this query and the results returned to the user.
To measure the accuracy, we consider the correctness and the completeness as  below:

\begin{center} 
\footnotesize
\vspace{-3mm}
$$Correctness = \frac{|R_{or} \cap R_{xs} |}{|R_{xs}|} \quad , \quad Completeness = \frac{|R_{or} \cap R_{xs} |}{|R_{or}|} $$
\vspace{-2mm}
\end{center}

where $R_{or}$ is the set of results returned by the search engine for the original query, and $R_{xs}$ the set of results returned to the user.
Both metrics are in $[0,1]$. The best accuracy is provided with a correctness and a completeness at $1$.
We used the same methodology as described in~\cite{petit2015peas} to conduct the experiment.

\subsection{System Metrics}
\label{ssec:SysMetric}

To evaluate the behavior of \name from a systems perspective, we consider the following metrics.
First, we measure the end-to-end latency which is the time spent to serve search results back to users once they send their queries.
Second, we measure the throughput (requests/second) to assess the capability of \name to properly act as proxy even with a growing number of nodes requesting the same proxy.

\section{Evaluation}
\label{sec:evaluation}

We now present the results in term of sensitivity, privacy and system performance obtained by \name under the experimental setup of the previous section. 
Our results show that \name efficiently detects sensitive queries while limiting the number of overprotected queries actually not sensitive. 
We also show that \name provides a slightly better protection than state-of-the-art competitors and drastically reduces the end-to-end latency without any impact on the accuracy.

\subsection{Privacy: Robustness Against Re-Identification Attack}
\label{ssec:EvalPrivacy}

This section evaluates the capacity of \name to protect the user privacy by 
measuring its robustness against an adversary conducting a re-identification attack.
Figure~\ref{fig:PrivacyEval-ReidentifAttack} depicts the re-identification rate for \SYS, \tor, \tmn, \goopir, \peas, and \xsearch with $k=7$.

Without query obfuscation (i.e., \tor), an adversary is ensured that every received query has been issued by a user.
Consequently, the challenge in this case consists to map each received query to preliminary information collected about users.
Results show that an adversary using past queries of users as prior knowledge 
is able to re-affiliate around $36\%$ of the new queries to their original users.
Interesting enough, result for \tor also represents the re-identification rate of \peas, \xsearch and \name with $k=0$.

Without unlinkability (i.e., \tmn and \goopir), the re-identification rate corresponds to retrieve the real queries from the fake ones.
Results show that the adversary is able to retrieve a large proportion of real queries, 45\% and 50\% for \tmn and \goopir, respectively. 
This high re-identification rate mainly comes to the fake query generation process which uses RSS feeds to build fake queries. 
If the content of these RSS is far from the interests of the user, the adversary can easily dissociate them.

Combining query obfuscation and unlinkability drastically drops the re-identification rate.
Indeed, the challenge for the adversary becomes harder and consists to retrieve both the identity of the user and the real queries from the fake ones.
Generate fake queries is challenging. 
These fake queries have to be indistinguishable from real ones.
By using real past queries as fake ones, \xsearch and \name provide a lower re-identification rate than \peas which generate fakes queries using a graph of co-occurence of terms built from past queries.
Finally, \name slightly reduces this re-identification rate compared to \xsearch (e.g., $6\%$ for \xsearch compared to $4\%$ for \name).
This difference comes from the obfuscation scheme of these solutions.
For \xsearch, the adversary receives a group of (k+1) queries and has to identify the real one among this group.
For \name, as the adversary receives individually each query whether it is a real query or a fake one, the re-identification process is much harder and creates more confusion for the adversary.

\begin{figure*}[tb]
	\begin{subfigure}{0.246\textwidth}
		\centering
		\includegraphics[scale=0.5,trim=0 0 0 0]{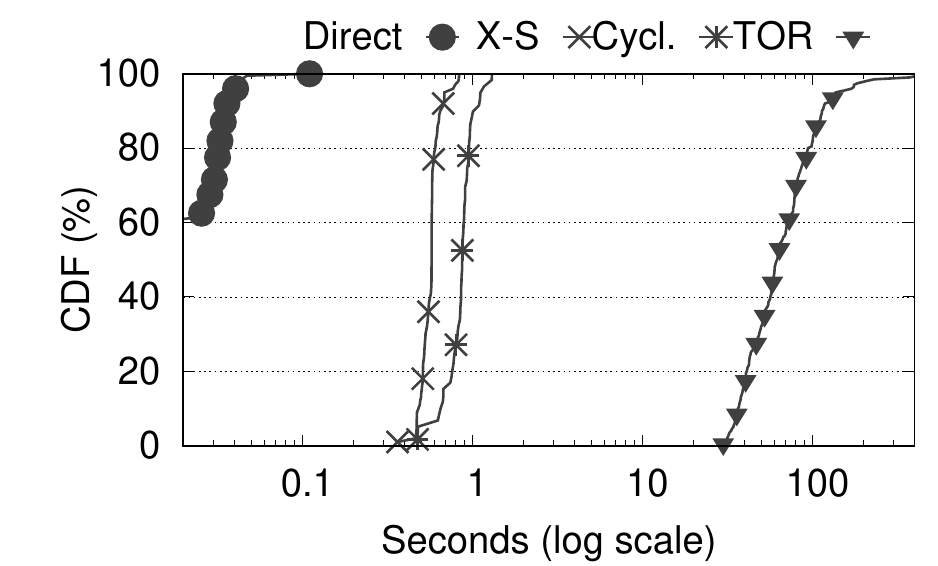}
		\caption{End-to-end delays for 200 queries, $k=3$.}
		\label{fig:endtoend-latency}
	\end{subfigure}
		\begin{subfigure}{0.246\textwidth}
		\centering
		\includegraphics[scale=0.5,trim=0 0 0 0]{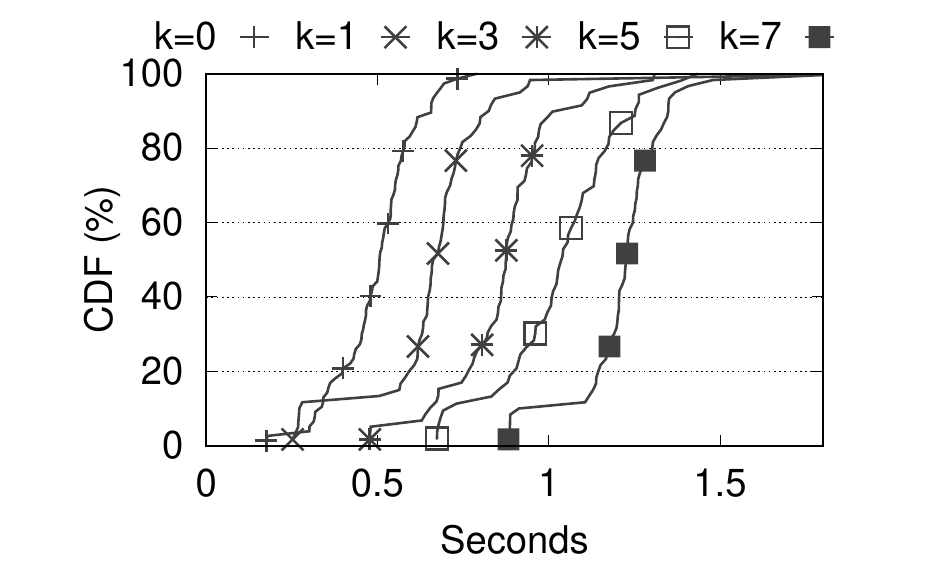}
		\caption{Impact of k on observed latency.}
		\label{fig:endtoend-latency-fakes}
	\end{subfigure}
		\begin{subfigure}{0.246\textwidth}
		\centering
		\includegraphics[scale=0.5,trim=0 0 0 0]{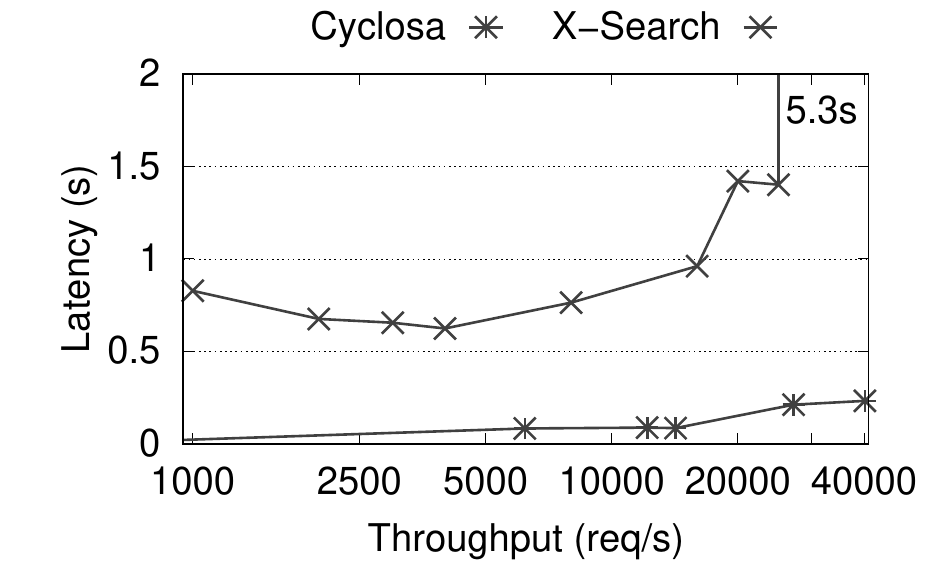}
		\caption{Throughput/Latency of \name and \xsearch.}
		\label{fig:tput_lat}
	\end{subfigure}
		\begin{subfigure}{0.246\textwidth}
		\centering
		\includegraphics[scale=0.5,trim=0 0 0 0]{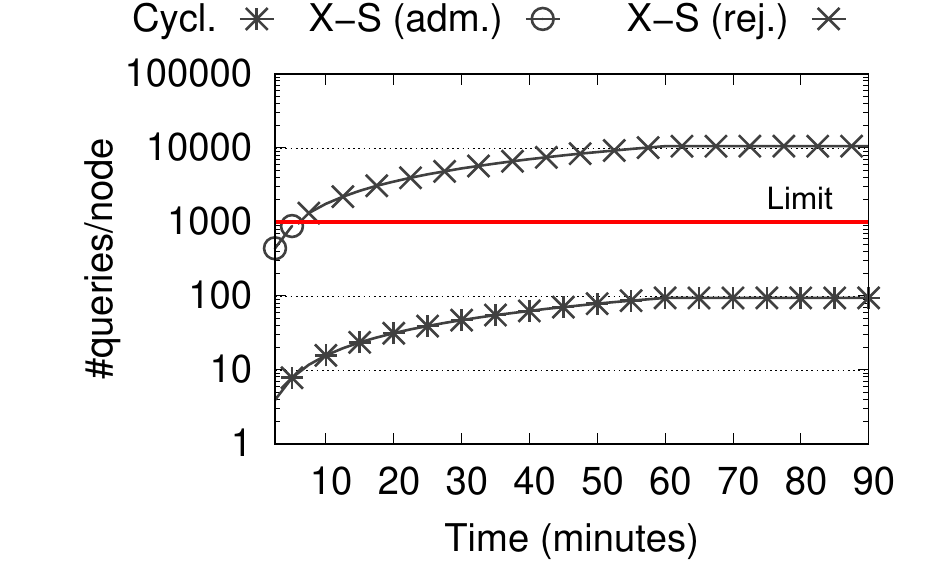}
		\caption{Query protection vs. users blocked by search engine}
		\label{fig:load}
	\end{subfigure}
	\caption{Multiple measurement results for \xsearch (X-S) and \name (Cycl.)}
\vspace{-2mm}
\end{figure*}

\subsection{Accuracy of Private Web Search}
\label{ssec:EvalAccuracy}
 
We evaluate the accuracy of \name, i.e.,~its ability to return to protected user queries the same answers from the search engine as the ones of non-protected queries.
Figure~\ref{fig:AccuracyEval-WebSearch} depicts the correctness and completeness of answers returned by \SYS, and by its competitors \tor, \tmn, \goopir, \peas, and \xsearch with $k=3$.
There are two sets of solutions.
\SYS as well as \tmn provide perfect accuracy, because either they do not apply obfuscation (e.g.,~\tor), or they differentially handle real and fake queries' responses.
The other solutions provide lower accuracy because they are not able to distinguish between answers of fakes queries or real query.
They try to extract the answer to the real query by filtering the union of answers to all (fake and real) queries, with an imperfect result.
Here, the precision reaches $65\%$ for a recall of $70\%$ for $k=3$.
These values decrease for a larger $k$ as reported in~\cite{benmokhtar:hal-01588883}.

\subsection{Adaptive Query Protection}

\name dynamically and adaptively protects queries according to their sensitivity.
Figure~\ref{fig:kvalues} reports the Cumulative Distribution Function (CDF) of the actual number of fake queries induced by \SYS to protect queries in our testing set when the maximum value of $k$ is defined at $7$.
Results show that $25\%$ of queries do not need fake queries, and $50\%$ of them use less than $3$ fake queries.
The sharp increase reported for $k=7$ corresponds to queries identified as highly sensitive, and consequently requiring the maximum protection level.
In the underlying workload, only 35\% of queries require that maximum number of fake queries.
In contrast, \xsearch would have generated, for each user query, that maximum number of fake queries .

\subsection{System Evaluation}
\label{ssec:PerformanceEval}
We begin by showing the observed end-to-end latency of the queries issued to the search engine by a client.
In this benchmark, we compare the results of \name against \xsearch and \tor.
We further include the measuraments achived without any protection and contacting directly the search engine. 
Figure~\ref{fig:endtoend-latency} presents our results.
On the x-axis (log-scale) we show the measured latencies, on the y-axis the respective CDF.
As expected, \tor is the slowest of them, with a median latency of 62.28 seconds.
We observe how both \name and \xsearch allow for sub-seconds results for the large majority of the queries, with a median of 0.876 and 0.577 seconds respectively.
We explore the impact of changing the number of issued fake queries in Figure~\ref{fig:endtoend-latency-fakes}.
We observe that by more than doubling the issued fake queries (from k=3 to k=7), the system still returns the results to the clients in less than 1.5 seconds in the worst case (median latency at 1.226 seconds).
These performances allow \name to offer a usable web browsing experience without negatively affecting the web publishers' revenue model~\cite{usability-web}.

We also evaluate the capacity for a \name node to substain very high rates of request/seconds.
Figure~\ref{fig:tput_lat} presents our results against \xsearch. 
We submit requests at increasingly high constant rates, and measure the latency to return a reply to the client from the next hop in the workflow chain (the \xsearch proxy or a \name relay), but without actually submitting the requests to the search engine.
\name is able to handle very high requests rates with sub-seconds response delays.
In our evaluation, we achieved a $40,000$ requests/sec with a 0.23s median response delay while \xsearch starts straggling at $30,000$ requests/sec.

Note that such rates, although possible, cannot be observed in practice without being immediately blocked by a smart search engine's malicious user detection system.
In our experiments, this happened very soon.
Although some services offer high read rates~\cite{searchap-kg}, users submit searches at rates orders of magnitude slower: the 100 most active users from the AOL dataset only submit 31.23~queries/hour.
For protecting such queries, \xsearch induces 10,500 req/hour among real and fake ones for $k=3$; and hence, it is eventually blocked by the search engine.
\SYS follows a more practical approach by spreading the load among the nodes with up to 94 req/hour per node for $k=3$, as shown in the simulation-based results presented in Figure~\ref{fig:load}.

\subsection{Accuracy of \SYS's Automatic Query Categorizer}
\label{ssec:AccuracyQueryCategorizer}

In this section we evaluate how \name is able to identify semantically sensitive queries.
As defined in Section~\ref{ssec:sensitivity}, the sensitivity of a query is evaluated over two dimensions 
capturing the linkability of the query and if it belongs to a topic defined as sensitive by the user.
The semantically sensitive query detection is based on both WordNet libraries and a trained LDA statistical model
(see Section~\ref{sssec:LexicalSemanticsBasedAnalysis}). Table~\ref{tab:detection} reports the precision and the recall of the detection of queries belonging in the sensitive topic related to sexuality.

Overall, most of queries related to the sensitive topic are detected, with a recall between $0.83$ for WordNet to $0.89$ for LDA.
Here, the precision can vary from $0.53$ to $0.86$ for WordNet, and when combining WordNet and LDA.
Precision captures the number of queries marked as sensitive according to the number of actually sensitive queries.
The closest to $1$, the better to avoid to overprotecting non-sensitive ones.
Results show a trade-off between precision and recall. Combining WordNet and LDA provides the better trade-off by identifying most of the sensitive queries while limiting the number of overprotected queries.

\begin{table}[!t]
\setlength{\tabcolsep}{15pt}
\center
\begin{tabular}{l c c}
 \toprule
 \textbf{Semantic tool} & \textbf{Precision} & \textbf{Recall} \\
  \midrule
  WordNet & 0.53 & 0.83 \\
  LDA & 0.84 & 0.89 \\
  WordNet + LDA & 0.86 & 0.85 \\
 \bottomrule
\end{tabular}
\caption{Detection of semantically sensitive queries}
\label{tab:detection}
\end{table}

 \section{Conclusion}
\label{sec:conclusion}

This paper presented \SYS, the first decentralized, private and accurate Web search solution that protects Web users against the risk of re-identification.
\SYS provides adaptive privacy protection, leveraging different query sensitivity levels by combining the analysis of user query linkability and query semantic.
\SYS follows a fully decentralized architecture for higher scalability, and is based on Intel SGX trusted execution environments for preventing from user data leakage between the nodes of the decentralized architecture.
\SYS reaches perfect accuracy of results of protected Web queries in comparison with non-protected queries.

Our implementation, evaluation results and comparison with state-of-the-art solutions show that \SYS is the most robust system against user re-identification, provides the most accurate results of Web search, in an efficient and scalable way.
Future work will investigate other datasets and workloads with different query sensitivity levels.
Although the protection solution presented in the paper addresses Web search, we believe that other application areas could be considered. 
\section*{Acknowledgment}

This work has received funding from the European Union’s Horizon 2020 research and innovation programme and was supported by the Swiss State Secretariat for Education, Research and Innovation under grant agreement No 690111 (SecureCloud). This work was also partially funded by French-German ANR-DFG project PRIMaTE (ANR-17-CE25-0017).

{
\printbibliography
}

\end{document}